\documentclass{appolb}
\usepackage{geometry}
\usepackage{graphicx}
\usepackage{float}
\usepackage[compress,numbers,square]{natbib}
\usepackage{hyperref}
\usepackage{graphicx}
\usepackage{soul}
\usepackage{blindtext}
\usepackage{colortbl}
\usepackage{xcolor}
\usepackage{multirow}
\usepackage{wrapfig}
\usepackage{multicol}
\usepackage{doi} 
\usepackage{wrapfig}

\usepackage[T1]{fontenc}
\usepackage{ae} 
\usepackage{aecompl}
\usepackage{tgbonum}

\newenvironment{Figure}
  {\par\medskip\noindent\minipage{\linewidth}}
  {\endminipage\par\medskip}

\begin{document}
\title{\textbf{Nanoparticle seeded glancing-angle deposition of tip-handle heterostructures}}

\author{Kai Trepka$^{a,b,}$\footnote{Corresponding author. kai.trepka@ucsf.edu}, Govind Bindra$^a$, Haley Langan$^a$, Jessica Lin$^a$, Kristina Linko$^a$, Henry Tsang$^a$, Nare Janvelyan$^a$, Fanny Hiebel$^a$, Ye Tao$^a$
\address{$^a$Rowland Institute at Harvard, Massachusetts, USA \\$^b$Medical Scientist Training Program, University of California San Francisco, California, USA}
}
\maketitle
\begin{abstract}
The controllable handling of an arbitrary single particle of matter with sub-$100$ nanometer (nm) dimensions is an essential but unsolved scientific challenge. We demonstrate nanoparticle-seeded glancing angle deposition using $10-100$ nm diameter nanoparticle seeds (Er$_2$O$_3$, Fe@C, and Fe). The products are nanoparticle-nanowire heterostructures composed of arbitrary nano-scale tips attached to micron-length nanowire handles. Optical micromanipulation of the micron-scale handles enables concurrent manipulation of the attached nanoscale particles of matter.
\end{abstract}

\pagestyle{plain}

\section{Introduction}

\begin{figure}[htb]
  \centering
  \includegraphics[width=0.5\linewidth]{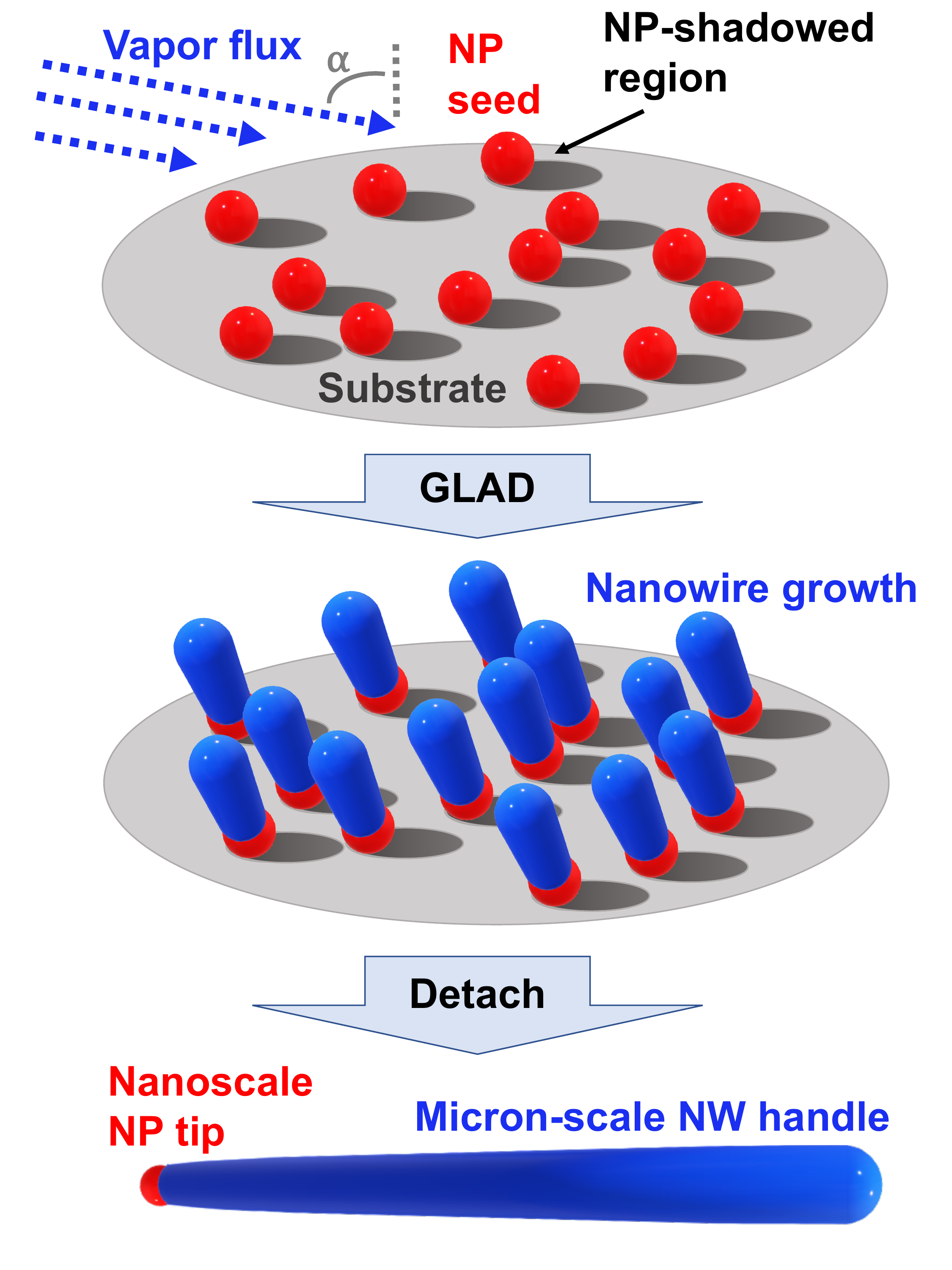}
  \caption{Nanoparticle-seeded GLAD. Top: GLAD of a source material (blue arrows) onto well-dispersed seeds (red spheres). Middle: nanoparticle-nanowire array on substrate. Bottom: individual heterostructure, detached via micromanipulator.}
  \label{fig:sch}
\end{figure}

\noindent Generalized manipulation of single sub-$100$ nm particles of matter is an outstanding technological challenge, with applications including functionalized tip fabrication for scanning probe microscopy (SPM) and intracellular sensing and perturbation \cite{mrfm,mrfm2,cell}. While tools such as transmission electron microscope \textit{in-situ} manipulation, atomic force microscope attachment, optical tweezers, and application-specific bottom-up nanofabrication efforts have driven scientific progress, none are simultaneously scalable and generalizable to arbitrary nanoparticles (NP) \cite{TEM,AFM,tweezers,application}.
\\ \\
\noindent Glancing angle deposition (GLAD) is a physical method of porous thin film growth where a source material is evaporated at an oblique angle with respect to a substrate. Spontaneous adsorption of vaporized atoms results in self-shadowing seeds that elongate into parallel nanocolumn arrays \cite{GLAD2}. Here, we demonstrate NP-seeded GLAD to make detachable and manipulable tip-handle heterostructures. Arbitrary tip NPs are used as the seeds for nanowire (NW) handle growth, followed by micromanipulator-mediated detachment of individual tip-handle heterostructures (Fig. \ref{fig:sch}).

\section{Methods}
\noindent Commercial Er$_2$O$_3$ and Fe@C NP with lateral dimensions $20-100$ nm were dispersed onto Si substrates via spin coating. $10$ nm Fe NP were synthesized as described previously \cite{Fe}, followed by dispersion onto substrate. Glancing angle deposition of nanowire handles was performed on the nanoparticle-seeded substrates. Using thermal GLAD, SiO and Ge handles were grown on Fe@C- and Er$_2$O$_3$-seeded substrates, respectively. Using e-beam GLAD, Ti and FeCo handles were grown on Fe-seeded substrates \cite{FeCo}. All products were characterized via scanning electron microscopy (SEM). See Tab. S1 and S2 for material and process details.

\section{Results and conclusions}

\begin{figure}[htb]
  \centering
  \includegraphics[width=0.5\linewidth]{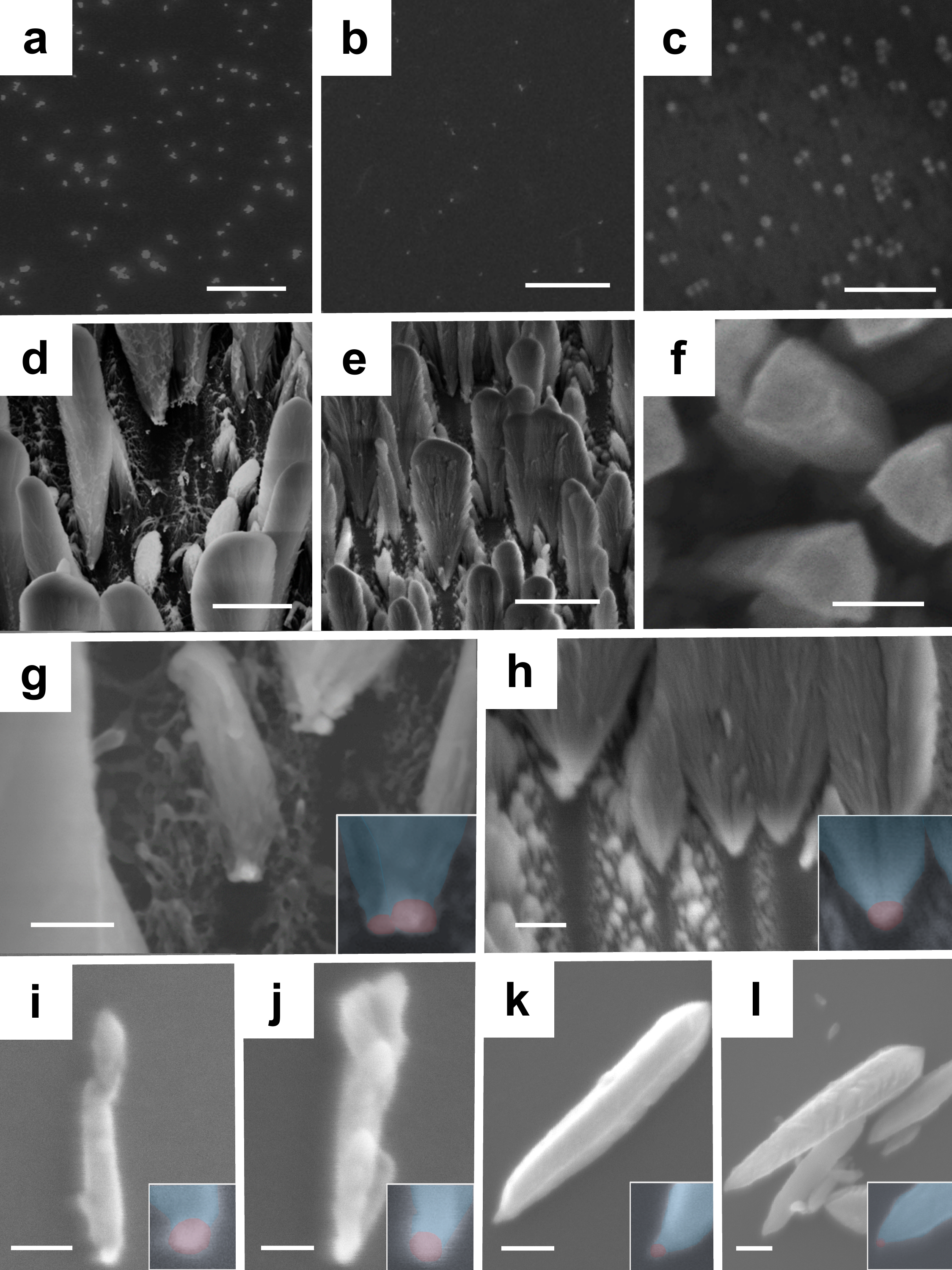}
  \caption{Scanning electron micrographs of glancing angle-deposited tip-handle heterostructures. Nanoparticle dispersion on Si substrates: (a) Er$_2$O$_3$, (b) Fe@C, (c) Fe. Heterostructure arrays: (d) Er$_2$O$_3$ NP-Ge NW, (e) Fe@C NP-SiO NW, (f) Fe NP-FeCo NW. Single heterostructures: (g) Er$_2$O$_3$ NP-Ge NW, (h) Fe@C NP-SiO NW, (i,j) Fe NP-Ti NW, (k,l) Single Fe NP-FeCo NW. Scale bars (nm): 1000 (a,b,d,e), 200 (h), 100 (c,f,g,k,l), 50 (i,j). Insets are magnified (2x) and colorized, with red tips and blue handles. See Tab. S3 for quantification.}
  \label{fgr:fig1}
\end{figure}

\noindent Generalized nanoparticle tip-nanowire handle heterostructures were grown using glancing angle deposition. Er$_2$O$_3$, Fe@C, and Fe nanoparticles were dispersed onto Si substrates (Fig. \ref{fgr:fig1}a-c), followed by GLAD of Ge, SiO, FeCo, and Ti handles. The products are tip-handle heterostructures with varied chemical and magnetic properties, illustrating process generality across diverse tip and handle materials (Fig. \ref{fgr:fig1}d-f). The differing morphologies are likely due to differing growth methods, with thermal deposition resulting in more rounded nanowire handles (Fig. \ref{fgr:fig1}d,e), and e-beam deposition resulting in sharper structures visually consistent with greater crystallinity (Fig. \ref{fgr:fig1}f).
\\ \\
\noindent Nanoscale tips can be optically micromanipulated via microscale handles. Individual heterostructures contain a single NP tip attached to a microscale NW handle that spreads out from the tip (Fig. \ref{fgr:fig1}g-l). An optical micromanipulator was used to physically interact with individual handle structures and lift them off of the primary substrate. The heterostructures retain their morphology upon detachment from the growth substrate and placement on a secondary substrate, suggesting that the tip-handle adhesive force is qualitatively stronger than the tip-substrate adhesive force (Fig. \ref{fgr:fig1}g-l). 

\noindent The presented physical deposition method enables a versatile selection of tip-handle combinations, while established methods such as vapor-liquid-solid (VLS) growth of tip-handle heterostructures produce more uniform handles at the expense of generality \cite{vls}. Further, the presented method requires no heating of the nanoparticle dispersions, a process that could potentially denature surface groups and diminish particle properties.
\\ \\
\noindent While the variety of material combinations reported here suggests both chemical and dimensional generality of seeded GLAD, three important factors need to be considered. First, the desired tip nanoparticles must be well-dispersed on a growth substrate and thus the dispersion step may require special attention when considering aggregation-prone particles. Second, to enable detachment of the structures from the substrates, the substrate material is selected for its relatively low adhesion to the particles when compared to the particle-handle adhesion. In instances where the substrate adherence is too strong, depositing the particles on a soluble mask prior to deposition and dissolving the mask to detach the heterostructures post-deposition may be a viable option. Finally, the growth process itself imposes little
limitation on the handle material, since most inorganic compounds are amenable to physical evaporation and deposition.
\\ \\
\noindent The nanoparticle-handle construct provides advantages beyond traditional nanoparticle experimentation techniques including physicochemical characterization of individual particles, the use of individual particles in directed, local perturbation, and particle-based sensing with 3-dimensional access. The handle allows rapid and facile manipulation of individual particles under optical microscopy, and could be designed to allow electronic access to individual nanoparticles for chemical or sensor applications. The wire geometry allows direct product use as oscillators and facile attachment force-sensing cantilevers \cite{cantilever}. Because rare-earth oxides such as Er$_2$O$_3$ and Ho$_2$O$_3$ combine high saturation magnetization with a long dynamic range and an absence of hysteresis \cite{magnet,reo}, one future application
may use a rare-earth oxide-tipped heterostructure as a
resonant force sensor to improve the sensitivity, quantitativeness, and resolution of magnetic SPM \cite{mrfm}.
\\ \\
\noindent In conclusion, we have demonstrated a proof of concept for handle-mediated manipulation of individual sub-100 nm nanoparticles with varied dimensions and compositions. GLAD
imposes minimal technical restrictions in terms of material
choice, putting the final scientific purpose at the center of the design process.

\section{Acknowledgements}
\noindent This work was supported by the Rowland Institute at Harvard. Data are available on GitHub (https://github.com/trepkakai/glad-manipulation). The authors declare no competing interests.

\bibliographystyle{unsrtnat}

\bibliography{manuscript}

\newpage

\section{Supplemental information}

\subsection{Nanoparticle synthesis and dispersion
}

\noindent Er2O3 and Fe@C NP with lateral dimensions 20-100 nm were dispersed onto Si substrates via spin coating. 40-50 nm Er2O3 NPs were diluted in Milli-Q deionized water (10 mg/ml), followed by 24 hr of bath sonication, 24 hr of sedimentation, and extraction of the supernatant as the solution. 20-100 nm Fe@C NPs were diluted in DCB (4 mg/ml) and probe sonicated for 5 minutes (10 seconds on, 10 seconds off, power level 10). Si substrates were prepared as described in prior work. Spin coating for 30 seconds (3000 rpm, 1000 rpm/s ramp) dispersed the Er2O3 and Fe@C NP solutions onto the substrate. 10 nm Fe NP were synthesized via homogeneous nucleation with 100 mM Fe(CO)5 precursor as described previously, followed by dilution (10x) of the reaction mixture in hexane, pipetting onto substrate, and N2 blow-drying. 
\\ \\ \\ \\

\begin{Figure}{Table S1: Key resources table}
\small
\centering
\begin{tabular}{|l|l|l|}
\hline
\rowcolor[HTML]{E7E6E6} 
\textbf{RESOURCE}            & \textbf{SOURCE}      & \textbf{IDENTIFIER}                                                                                        \\ \hline
\rowcolor[HTML]{E7E6E6} 
\multicolumn{3}{|l|}{\cellcolor[HTML]{E7E6E6}Instrumentation}                                                                                                    \\ \hline
Thermal evaporator           & Denton               & DV-502                                                                                                     \\ \hline
E-beam evaporator            & Plassys              & Plassys II                                                                                                 \\ \hline
Scanning electron microscope & Zeiss                & Supra-55 VP                                                                                                \\ \hline
Bath sonicator               & Branson Ultrasonics  & Branson 2800                                                                                               \\ \hline
Probe sonicator              & QSonica              & Q125                                                                                                       \\ \hline
\rowcolor[HTML]{E7E6E6} 
\multicolumn{3}{|l|}{\cellcolor[HTML]{E7E6E6}Tip and tip preparation materials}                                                                                  \\ \hline
Er2O3 nanoparticles          & SSNano               & 3010DX                                                                                                     \\ \hline
Fe@C nanoparticles           & Sigma-Aldrich        & 746827                                                                                                     \\ \hline
Fe nanoparticles             & Previously described & \begin{tabular}[c]{@{}l@{}}DOI:10.1021/jp0351831;\\ homogeneous nucleation method\end{tabular} \\ \hline
\rowcolor[HTML]{E7E6E6} 
\multicolumn{3}{|l|}{\cellcolor[HTML]{E7E6E6}Handle and handle preparation materials}                                                                            \\ \hline
SiO pellets                  & Lesker               & EVMSIO-1065B                                                                                               \\ \hline
Ge pellets                   & Lesker               & EVMGE-1038B                                                                                                \\ \hline
Ti metal                     & Lesker               & EVMTI45QXQD                                                                                                \\ \hline
FeCo                         & Previously described & DOI:10.1038/s41467-017-02519-8                                                                 \\ \hline
\rowcolor[HTML]{E7E6E6} 
\multicolumn{3}{|l|}{\cellcolor[HTML]{E7E6E6}Substrate and substrate preparation materials}                                                                      \\ \hline
Silicon wafers               & Previously described &                                                                                                            \\ \hline
\end{tabular}
\end{Figure}

\newpage
\subsection{Glancing angle deposition}

\noindent Subsequent glancing angle deposition of nanowire handles was performed on the nanoparticle-seeded substrates. GLAD of SiO was carried out on Fe@C-seeded substrates, Ge on Er2O3-seeded substrates, and FeCo and Ti on Fe-seeded substrates. GLAD of SiO and Ge was carried out in a thermal evaporator with azimuthal angle $\alpha$ = 85° and substrate cooling (T = 100 K), with source temperature adjusted to maintain a deposition rate of 3-6 Å/s as measured by quartz crystal microbalance. GLAD of Ti and FeCo handles was performed in an e-beam evaporator, with $\alpha$ = 85°, a substrate rotation rate of 15 rpm, and evaporation rate adjusted to maintain a deposition rate of 5 Å/s. All products were characterized via scanning electron microscopy.
\\ \\ \\ \\

\begin{Figure}{Table S2: Glancing angle deposition process parameters}
\centering
\small
\begin{tabular}{|l|l|l|l|l|}
\hline
\textbf{\begin{tabular}[c]{@{}l@{}}Tip-handle \\     heterostructure\end{tabular}}    & \textbf{\begin{tabular}[c]{@{}l@{}}NP     \\ dilution\end{tabular}}                                                                                            & \textbf{\begin{tabular}[c]{@{}l@{}}NP dispersion in     \\ solution\end{tabular}}                                                                    & \textbf{\begin{tabular}[c]{@{}l@{}}NP dispersion on     \\ substrate\end{tabular}}                                                                                                     & \textbf{\begin{tabular}[c]{@{}l@{}}GLAD     \\ parameters\end{tabular}}                                                                                                                                                                                                                  \\ \hline
\begin{tabular}[c]{@{}l@{}}Er2O3 NP-Ge     \\ NW\end{tabular}                         & \begin{tabular}[c]{@{}l@{}}Dilution of     \\ Commercial    \\ powder in     \\ Milli-Q     \\ deionized     \\ water (10     \\ mg/ml)\end{tabular} & \begin{tabular}[c]{@{}l@{}}24 hrs bath     \\ sonication, 24 hr     \\ sedimentation,     \\ extraction of     \\ supernatant\end{tabular}     & \begin{tabular}[c]{@{}l@{}}Spincoating of 200 \\     $\mu$l solution on     \\ $0.7x0.7$" Si    \\ substrates for 30     \\ seconds (3000 rpm,    \\ 1000 rpm/s ramp)\end{tabular} & \begin{tabular}[c]{@{}l@{}}Thermal     \\ evaporator,     \\ azimunthal     \\ angle $\alpha$ = \\ 85$^{\circ}$,     substrate     \\ cooling (T =     \\ 100 K), source     \\ temperature     \\ adjusted to     \\ maintain    \\ deposition rate     \\ of 3-6 Å/s\end{tabular} \\ \hline
\begin{tabular}[c]{@{}l@{}}Fe@C NP-SiO     \\ NW\end{tabular}                         & \begin{tabular}[c]{@{}l@{}}Dilution of     \\ commercial     \\ powder in     \\ DCB (4     \\ mg/ml)\end{tabular}                                       & \begin{tabular}[c]{@{}l@{}}5 minutes probe     \\ sonication (10     \\ seconds on, 10     \\ seconds off, power     \\ level 10)\end{tabular} & \begin{tabular}[c]{@{}l@{}}Spincoating of $200$     \\ $\mu$l solution on     \\ $0.7x0.7$" Si     \\ substrates for $30 $    \\ seconds ($3000$ rpm,     \\ $1000$ rpm/s ramp)\end{tabular} & \begin{tabular}[c]{@{}l@{}}Thermal     \\ evaporator, $\alpha$ =     \\ 85$^{\circ}$, substrate     \\ cooling (T =     \\ $100$ K), source     \\ temperature     \\ adjusted to     \\ maintain     \\ deposition rate     \\ of $3-6$ Å/s\end{tabular}                                    \\ \hline
\begin{tabular}[c]{@{}l@{}}Fe NP-Ti NW     \\ and Fe NP-    \\ FeCo NW\end{tabular} & \begin{tabular}[c]{@{}l@{}}Dilution of     \\ reaction     \\ mixture in     \\ hexane     \\ (1:10)\end{tabular}                                        & N/A                                                                                                                                                    & \begin{tabular}[c]{@{}l@{}}Pipetting onto     \\ $0.7x0.7$" Si     \\ substrates, N2 blow-    \\ drying\end{tabular}                                                                 & \begin{tabular}[c]{@{}l@{}}e-beam     \\ evaporator, $\alpha$ =     \\ 85°, substrate     \\ rotation 15     \\ rpm,     \\ evaporation \   \\ rate adjusted to     \\ maintain a     \\ deposition rate    \\ of 5 Å/s\end{tabular}                                           \\ \hline
\end{tabular}
\end{Figure}

\newpage
\subsection{Nanoparticle seed density vs heterostructure density
}

\noindent One metric of yield is the seed particle density vs as-grown nanowire density, a measure of how many nanoparticle seeds end up attached to the resulting grown nanowire handles for manipulation post-GLAD. Qualitatively, all nanowire handles appear attached to seeds, while not all nanoparticle seeds are attached to handles. For the micrographs shown in Figure 2, the density of distinct particles, density of distinct as-grown heterostructures, and ratio between the two are reported (Table S3). ImageJ was used for the counting, and any continuous aggregate of particles were counted as a single seed. 
\\ \\ \\ \\
\begin{Figure}{Table S3: Nanoparticle seed vs heterostructure density for Figure 2}
\centering
\small
\begin{tabular}{|l|l|l|l|l|}
\hline
\textbf{\begin{tabular}[c]{@{}l@{}}Tip \\ material\end{tabular}} & \textbf{Figures} & \textbf{\begin{tabular}[c]{@{}l@{}}Seed particle tip \\ density, $n_t$ ($\mu$m$^{-2}$)\end{tabular}} & \textbf{\begin{tabular}[c]{@{}l@{}}Nanowire handle \\ density, $n_h$ ($\mu$m$^{-2}$)\end{tabular}} & \textbf{\begin{tabular}[c]{@{}l@{}}Fraction of \\ particles with \\ handles, $n_t/n_h$\end{tabular}} \\ \hline
Er2O3                                                            & 2a, 2d           & 6.3                                                                                      & 1.0                                                                                    & 0.16                                                                                             \\ \hline
Fe@C                                                             & 2b, 2e           & 1.8                                                                                      & 0.95                                                                                   & 0.53                                                                                             \\ \hline
Fe                                                               & 2c, 2f           & 600                                                                                      & 27                                                                                     & 0.046                                                                                            \\ \hline
\end{tabular}
\end{Figure}

\end{document}